\documentstyle[12pt]{article}
\begin{document}
\title{Causal heat conduction contravening the fading memory paradigm}
\author{L. Herrera$^1$\thanks{e-mail: lherrera@usal.es}\\
{$^1$Instituto Universitario de F\'isica
Fundamental y Matem\'aticas},\\ {Universidad de Salamanca, Salamanca 37007, Spain}}
\maketitle

\abstract{We propose a causal heat conduction model based on a heat kernel violating  the fading memory paradigm. The resulting transport equation produces  an  equation for the temperature. The model is applied to the discussion of two important issues such as the thermohaline convection and the nuclear burning (in)stability. In both cases the behaviour of the system appears to be strongly dependent on the transport equation assumed, bringing out the effects of  our specific kernel on the final description of these problems. A possible relativistic version of the obtained transport equation is presented.} 

\section{Introduction}
In the study of dissipative processes,   the order of magnitude of relevant  time scales of the system (in particular the thermal relaxation time) play a major role, due to the fact that  the interplay between this latter time scale and the time scale of observation, critically affects the obtained  pattern of evolution. This applies not only for  dissipative processes
out of  the steady--state regime, but also when the observation time is of the
order of (or shorter than)  the  characteristic time of the system under
consideration. Besides,  as it has been stressed before \cite{Ref1}, for time scales larger than the relaxation time, transient
phenomena affect the future of the system, even for time scales much
larger than the relaxation time.

Starting with the original Maxwell works, dissipative processes were initially studied by means of parabolic theories. These have been  proved very useful especially in the steady--state regime \cite{Ref2}.

In the context of these theories,  the  well known  classical  Maxwell--Fourier law for heat current  reads
\\
\begin{equation}
\vec{q}  = - \kappa \vec{\nabla}T \, ,
\label{q}
\end{equation}
\\
where $\kappa$, $T$ and $\vec q$ denote the heat conductivity of the fluid, the temperature and the heat flux vector respectively. Using the continuity equation
\begin{equation}
\frac{\partial u}{\partial  t}=-{\vec \nabla}.{\vec q},\label{ce1}
\end{equation}
 and the constitutive equation for the internal energy $u$
\begin{equation}
du=\gamma dT ,
\label{ene}
\end{equation}
 we obtain the parabolic equation for temperature (diffusion equation),
\\
\begin{equation}
\quad\frac{\partial T}{\partial t} = \chi \nabla^2T ,
\label{dif}
\end{equation}
\\
(where $\chi\equiv\frac{\kappa}{\gamma}$, and $\gamma$ are the diffusivity,  and 
heat capacity per volume, respectively), which, as is well known from the theory of parabolic differential equation, 
does not describe a causal  propagation of perturbations. In other words,  perturbations propagate with infinite
speed. 

This lack of causality is at the origin of alternative  proposals    based on    hyperbolic
theories of dissipation  \cite{Ref3,Ref4}, describing the propagation of perturbations with finite speed. 

These causal theories  include the dissipative fluxes as field variables, are supported  by statistical fluctuation theory and kinetic theory of gases, and  furthermore conform, in a high degree,  with experiments \cite{Ref5}.

In these theories a fundamental concept is that of  the relaxation time $\tau$ of the
corresponding  dissipative process. This positive--definite quantity measures  the time required  by the system to return
 to the steady state,  once  it has been abandoned. It must be emphasized that although it may be sometimes
connected to the mean collision time $t_{c}$ of the  particles responsible
of  the dissipative process, it should  not be identified with it.
Indeed, not only are these two time scales different, but still worse, there is  no general formula linking
$\tau$ and $t_{c}$.

 The origin of the pathological behavior exhibited by (\ref{q}) stems from  the fact that the Maxwell--Fourier law,  assuming  explicitly the vanishing of  the relaxation time,
implies that the heat flow starts (vanishes) simultaneously with
the appearance (disappearance) of a temperature gradient. 

However, even though $\tau$ is very small for many processes
such  as phonon--electron, and phonon--phonon interaction
at room temperature, (${\cal O}(10^{-11})$ and ${\cal O}(10^{-13})$ seconds,
respectively \cite{Ref6}), neglecting it produces the acausality mentioned above and lead to unphysical predictions 
 as for example in superfluid Helium
\cite{Ref7,Ref8}, and degenerate stars where thermal conduction is dominated
by electrons.

In order to overcome this problem a generalization of the Maxwell--Fourier law was proposed by Cattaneo \cite{co,Ref9} and (independently) Vernotte \cite{Ref10}, by assuming a non--vanishing thermal relaxation time. The corresponding equation reads

\begin{equation}
\tau \frac{\partial \vec q}{\partial t} + \vec q = - \kappa \vec \nabla T.
\label{Cattaneo}
\end{equation}

This equation (known as Cattaneo-Vernotte's equation) leads to a
hyperbolic equation for the temperature (telegraph equation)
\\

\begin{equation}
\tau \frac{\partial^2 T}{\partial t^2 } + \frac{\partial T}{\partial t}
=  \chi \nabla^{2} T \, ,
\label{telegraph}
\end{equation}
\\
which describes the propagation of thermal signals with a finite speed
\\
\begin{equation}
v = \sqrt{\chi/ \tau}.
\label{velocity}
\end{equation}
\\

From all the above, it should be already obvious that the relaxation time cannot be neglected during the  transient regimes. Furthermore,   it is also obvious that for time scales of the order of (or smaller than)  the relaxation time,
neglecting the relaxation time  is equivalent to
disregard the whole process under consideration.

In the past it has been argued that  hyperbolic theories may not be necessary after all. The rationale behind this (wrong) conclusion is that  when  relaxation times are 
comparable  to the characteristic time of the system,  it is out of the
hydrodynamic regime.  Such argument is fallacious, and goes as follows: in the hydrodynamic regime (which we assume)
the ratio between the  mean free path of ${\bf fluid \quad particles}$  and
the characteristic length of the system must be lower that
unity, otherwise the regime becomes Knudsen's. In this latter case the material cannot longer be considered as  a fluid, in the usual sense. The fact that  sometimes (though not always), $t_c$  and $\tau$ may be of the same  order of magnitude, might lead to the erroneous conclusion that ``large'' relaxation times implies that the system is no longer in the hydrodynamic regime.

But that conclusion  is valid only in the case when  ${\bf fluid \quad particles}$ (the ones  making up the
fluid) and the ones that transport the heat are the same. However this is (almost)
never the case. For example, for a neutron star, $\tau$ is of the
order of the scattering time between electrons (which carry the
heat) , nevertheless we consider that the neutron
star is  formed by a Fermi fluid of degenerate neutrons. The same
can be said   for the second sound in superfluid Helium and solids, and
for almost any ordinary fluid. To summarize,  the hydrodynamic regime
refers to ${\bf fluid \quad particles}$, which  not necessarily (and as a matter of fact,
almost never) are responsible for the heat conduction.  Accordingly, large relaxation times (large
mean free paths of particles involved in heat transport) do not imply a
violation of the  hydrodynamic regime (this fact is often overlooked, see
\cite{Santos}
 for a more detailed discussion on this point).
 
Finally, it is worth mentioning that problems of the kind we analyze here, have recently attracted the attention of  researchers working on fractional calculus (see \cite{fc1, fc2}
 and references therein). Thus, it would be interesting to find out what new insights, on the problem under consideration here,  could be obtained by means of such methods.

\section{Thermal memory and integral representation of the heat flux vector}
It is useful  to write heat transport equation such as 
 (\ref{Cattaneo}) in an integral form. The idea behind such an approach comes from the theory of non--linear materials for which the stress at a point at any time is determined by  deformation gradients not only at that instant, but also at previous instants (see \cite{Ref11,Ref12,Ref13,Ref14}).
 
 For dissipative processes, integral representations of heat fluxes appear in  early works of Coleman, Gurtin and collaborators \cite{Ref15,Ref16,Ref17}.

Thus, solving (\ref{Cattaneo}) for ${\vec q}$ we obtain
\begin{equation}
\vec q = - \frac{\kappa}{\tau} \int_{-\infty}^{t}{\exp{\left[-
\frac{(t-t')}{\tau}\right]} \cdot \vec \nabla T(\vec x,t') dt'}.
\label{catanneointegral}
\end{equation}

Obviously, taking the $t$-defivative of (\ref{catanneointegral}) we get (\ref{Cattaneo}). 

The above equation is a particular case of the most general expression
\begin{equation}
\vec q = - \int^{t}_{-\infty}{Q(t-t') \vec \nabla T(\vec x,t') dt'},
\label{generalintegral}
\end{equation}
where the kernel  $Q$ describes the thermal memory of the material, its role consisting in distributing the relevance of   temperature gradients at different moments in the past. Also, it should be mentioned that in the expression above (as well as in all integrals expressions below) the integral domain selected is $(-\infty, t]$, instead of $[0, t]$, doing so we are implicitly neglecting the initial effects.

Thus assuming 
\begin{equation}
 Q = \kappa \delta(t-t'),
\label{kmf}
\end{equation}
we obtain
 \begin{equation}
  \vec q =
- \kappa \vec \nabla T \quad {\rm (Maxwell-Fourier)}.
\label{kmf1}
\end{equation}
In other words the Maxwell--Fourier law corresponds to a zero--memory material for which the only relevant
temperature gradient is the ``last" one, i.e., the one simultaneous with
the appearance of $\vec{q}$. 

On the other hand if we assume
\begin{equation}
 Q = \beta= {\rm constant},
 \label{wave}
 \end{equation}
we obtain
\begin{equation}
   \frac{
\partial^2 T}{\partial t^2} = \frac{\beta}{\gamma} \nabla^{2} T.,\label{wave1}
\end{equation}
where $\beta$ is a constant with units $\frac{[\kappa]}{[t]}$.
The above case corresponds to a material with   infinite memory, leading to an undamped wave with velocity  $v=\sqrt{\frac{\beta}{\gamma} }$. 

This particular case  describes propagation of thermal waves without attenuation, which is physically  objectionable. Furthermore the integral (\ref{generalintegral}) would diverge. To avoid this drawback, and to assure convergence it might be convenient, for this particular case, to write instead of (\ref{generalintegral}),
\begin{equation}
\vec q(x,t) -\vec q(x,t=0)= - \int^{t}_{0}{Q(t-t') \vec \nabla T(\vec x,t') dt'}.
\label{generalintegralbis}
\end{equation}
However, for the applications analyzed here, the convergence problem is not relevant and therefore we shall use (\ref{generalintegral}).

In this context, the
Cattaneo-Vernotte equation appears as compromise between the two extreme cases considered above, and  for which all temperature gradients
contribute to $\vec{q}$, but their relevance diminishes as  we go farther to the past, as it is apparent from its corresponding kernel:

\begin{equation}
 Q=\frac{\kappa}{\tau}e^{-(t-t')/\tau}.
\label{1qc}
\end{equation}

Thus, we can say that the temperature gradients in the ``neighborhood'' of the time  $t$ provide the main contribution for the heat flux vector appearing in $t+\tau$. Then, assuming that $\tau$ is sufficiently small, we may expand $\vec q$ around $t$ in power series of $\tau$ and keep only linear terms, whereas  the integral in (\ref{catanneointegral}) may be written approximately as $-\kappa \vec \nabla T(\vec x,t)$. Combining these two results, equation (\ref{catanneointegral}) becomes (\ref{Cattaneo}).

The  decreasing in the relevance of older temperature gradients as compared with the newer ones is referred  to as the ``fading memory'' paradigm, in the literature. Once again, this basic assumption  was introduced initially in the general theory of materials to express the idea that the recent history of deformation should have a greater effect than the remote one, on the present value of the stress \cite{Ref14,Ref18,Ref19}. The extension of  this assumption to the heat kernel \cite{Ref15,Ref16,Ref17,nun, li2} is straightfroward, and  seems to  fit  intuitively well with ``common sense''. 
Thus the kernels corresponding to the Cattaneo and the Maxwell--Fourier laws, satisfy the ``fading memory'' paradigm, whereas the kernel (\ref{wave}) does not.

However, no matter how intuitively acceptable the idea expressed by the ``fading memory'' paradigm may be, the fact remains that there is not a compelling reason to exclude  heat kernels not complying with   this paradigm for some  possible physical situations, and therefore we are perfectly legitimated to explore the possible existence  of heat kernels  not satisfying  the ``fading memory'' paradigm, and to analyze their potential applications to different physical scenarios. The next sections  are  devoted  to this endeavour.

\section{Violating the fading memory paradigm}

We shall now consider the possibility that, for reasons that will be discussed latter, the material under consideration is such that the heat flux vector depends stronger on the older temperature gradients than on the newer ones. In other words we shall consider a material whose thermal memory behaves all the opposite as expected from the fading memory paradigm . The resulting model  will be applied to two important physical issues, namely, the thermohaline convection and the (in)stability of nuclear burning.
\subsection{The kernel and the equation for the temperature}
Thus, let us consider a thermal kernel of the form:
\begin{equation}
 Q=c[1-e^{-(t-t')/\tau}],
\label{1nf}
\end{equation}
where $c$ is a constant with units $\frac{[\kappa]}{[t]}$. For simplicity we choose $c\equiv\frac{\kappa}{\tau}$.

Feeding back this kernel into (\ref{generalintegral}),  we obtain  for the heat flux  
\begin{equation}
 \vec q= - \frac{\kappa}{\tau}\int^{t}_{-\infty} \left[1-e^{-(t-t')/\tau}\right]\vec \nabla T(\vec x,t') dt'.
\label{Cattaneo1bonf}
\end{equation}

It is clear from the above expression that the oldest  temperature gradients have more influence on $\vec q$, that the present one. More so, the strongest influence comes from the gradient at infinite past  ($t^\prime\rightarrow - \infty$), whereas the gradient at $t=t^\prime$ (the present time) is irrelevant, thereby  contradicting the ``fading memory paradigm''. 

Next, using the Leibniz rule  it is  a simple matter to  find from  (\ref{Cattaneo1bonf}),

\begin{equation}
\frac{\partial \vec q}{\partial t} = - \frac{\kappa}{\tau^2} \int^{t}_{-\infty} e^{-(t-t')/\tau}\vec \nabla T(\vec x,t') dt',
\label{Cattaneo1onf}
\end{equation}
producing
\begin{equation}
\tau \frac{\partial \vec q}{\partial t} + \vec q = - \frac{\kappa}{\tau}  \int^{t}_{-\infty}\vec \nabla T dt'.
\label{Cattaneo1nf}
\end{equation}

The physical meaning of this last equation, becomes intelligible when we recall that in our model, unlike the Cattaneo--Vernotte equation,  the temperature gradients in the neighborhood of $t$ are irrelevant, and the most important contributions to $\vec q$ come from the remote past  ($t^\prime\rightarrow - \infty$).

From (\ref{Cattaneo1nf}), different equations for the temperature may be obtained, depending on the continuity equation to be assumed. This is a sensitive issue, since, in general, the continuity equation depends on the transport law (see \cite{chen, li} for a discussion on this point).Therefore, for any specific application of the resulting equation for the temperature, this issue must be handled with some care. Nevertheless, in the applications to be considered in the following sections, the explicit equation for the temperature wont be necessary.

 As an example,  for simplicity,  we shall  consider here the standard case, which might not be compatible with the proposed transport law. 
Thus, combining (\ref{Cattaneo1nf}) with the continuity equation (\ref{ce1}) for the internal energy $u$, we obtain
\begin{equation}
\frac{\partial^2 T}{\partial t^2 } + \frac{1}{\tau}\frac{\partial T}{\partial t}
= \frac{\kappa}{\tau^2\gamma}\int^{t}_{-\infty}\vec \nabla^2 T dt',
\label{telegraphnf}
\end{equation}
 or, taking the $t$-derivative of the above equation, we obtain the third order equation:

\begin{equation}
\frac{\partial ^3 T}{\partial t^3}+\frac{1}{\tau}\frac{\partial ^2 T}{\partial t^2}=\frac{\kappa}{\tau^2 \gamma} \vec \nabla^2 T.
\label{2}
\end{equation}

\subsection{Applications}
In the past, different  hyperbolic transport equations have been used to  study convection processes \cite{apl1,apl4,apl6,apl9,apl10,apl11,apl12}, and in general a  variety of  different interesting physical phenomena \cite{apl3,apl5,apl7,apl8,apl13}. In what follows we shall consider  two important problems involving dissipative processes, adopting our kernel (\ref{1nf}).

\subsubsection{Thermohaline instability}

Thermohaline convection may be observed in oceans whenever  a layer of warm salt water is
above a layer of fresh cold water. As far as the salt water is warm enough as to  reduce its
specific weight to below that of the fresh water, the
system is dynamically stable. Thus, if a blob of the upper layer is pushed
downward, buoyancy will push it back. 
However, the cooling of the warm salt water  leads to an increasing  of  its
density, decreasing thereby   the buoyancy, producing eventually  the sinking of  small blobs of salty
water (the so called ``salt fingers''). The interesting point is that instabilities of this kind can
also occur in stars, under a variety of circumstances \cite{KW}.

This kind of secular instability is controlled by the
heat leakage of the blob,  and therefore the sinking velocity of the blob critically depends on the heat transport equation used to calculate it. 

The general expression for the sinking velocity of the convective blob  reads (see \cite{KW} for details):
\begin{equation}
V_{sink.}=-\frac{H_p}{( \nabla_{ad}-\nabla)\tau_d}\frac{DT}{T}
\label{thm}
\end{equation}
where $\tau_d$ denotes  the thermal adjustment time, and $DT$ is the difference of the temperature between the convective blob and the surroundings. $H_P$ is the scale height of pressure,  defined as
\begin{equation}
H_P=-P\frac{dr}{dP},
\label{hp}
\end{equation}
where $P$ is the pressure and $r$ the spatial coordinate along the sinking direction of the blob. Finally, $\nabla$ measures the variation of the temperature with the pressure of the surroundings, whereas $\nabla_{ad}$ measures the variation of the temperature with pressure of the blob, at constant entropy ($E$), i.e.
\begin{equation}
\nabla\equiv \left(\frac{d\ln T}{d\ln P} \right),\quad     \nabla_{ad}\equiv \left(\frac{\partial \ln T}{\partial \ln P}\right)_E.
\label{nablas}
\end{equation}

As mentioned above, the critical point in the calculation of $V_{sink.}$ is the calculation of $DT(t)$.

Thus, let us consider a spherical convective blob with diameter $d$ and temperature $T_b$. Denoting by $T_s$ the temperature of the surrounding fluid, then $DT=T_b-T_s. $
Approximating the temperature gradient by $|\nabla T|  \approx 2DT/d$ we obtain for the energy loss $\lambda$ per unit of time from the whole surface $S$ of the blob: 
 
\begin{equation}
\lambda=S|\vec q|= \frac{2S}{d}\int^{t}_{-\infty}{Q(t-t')DT(t') dt'}.
\label{gthhm1}
\end{equation}
It is worth mentioning that the energy loss may be due to thermal conduction as well as radiation, for which we use the  diffusion approximation.

On the other hand we know from thermodynamics that   the rate by which the thermal
energy of the blob of volume $V $  is lost ($\lambda$), may be written as
\begin{equation}
\lambda=-\rho c_PV\frac{\partial T_b}{\partial t}\approx -\rho c_PV\frac{d DT}{dt},
\label{thin3}
\end{equation}
where $c_P$ and $\rho$ denote the specific heat at constant pressure and the mass density respectively.

Equating (\ref{gthhm1}) and (\ref{thin3}) we obtain:
\begin{equation}
\frac{d DT}{d t}=-\frac{1}{\kappa \tau_d}\int^{t}_{-\infty}{Q(t-t')DT(t') dt'},
\label{thin4}
\end{equation}
where $\tau_d\equiv\frac{\tau d^2 \rho c_P}{12k}$.

Therefore, from (\ref{thm}) different velocity profiles will be obtained depending on the specific description of the energy loss of the convective blob (i.e. the specific form of the kernel $Q(t-t')$).

Thus, if we use the kernel (\ref{kmf}) corresponding to the Maxwell--Fourier law,  (\ref{thin4})  becomes
\begin{equation}
\frac{d DT}{d t}=-\frac{1}{\tau_d}DT(t),
\label{thin4mf}
\end{equation}
whose  simple solution reads
\begin{equation}
 DT=DT(0)e^{-\frac{t}{\tau_d}},
\label{thin4mf}
\end{equation}
i.e. we have a monotonically cooling of the blob, with the e-folding time defined by the thermal adjustment time, as expected from purely physical considerations if we neglect the relaxation time.

However, if instead of using (\ref{kmf}), we use (\ref{wave}) then one obtains from (\ref{thin4}):
\begin{equation}
\frac{d DT}{d t}=-\frac{\beta}{ \kappa \tau_d}\int^{t}_{-\infty}{DT(t') dt'},
\label{thc}
\end{equation}
or
\begin{equation}
\frac{d ^2DT}{d t^2}+\frac{\beta}{ \kappa\tau_d}DT(t)=0,
\label{thcb}
\end{equation}
which describes an undamped  harmonic oscillation with frecuency
$\omega =\sqrt{\frac{\beta}{\kappa \tau_d}}$.

The thermohaline instability  using the kernel (\ref{1qc}) has been studied in \cite{hf}. In this case the equation (\ref{thin4}) for $DT$ becomes
\begin{equation}
\frac{d ^2DT}{d t^2}+\frac{1}{\tau}\frac{dDT}{d t}+\frac{1}{ \tau \tau_d}DT(t)=0,
\label{thcc}
\end{equation}
with a solution of the form:
\begin{equation}
DT=DT(0)e^{\alpha t} \cos{\omega t},
\label{thin4cu}
\end{equation}
describing a damped oscillation with frequency $\omega=\frac{1}{2\tau}\sqrt{\frac{4\tau-\tau_d}{\tau_d}}$ and damping factor $\alpha=-\frac{1}{2\tau}$.
Thus, as it sinks the convection blob oscillates, with a decreasing amplitude. The physical (observational) consequences of this behaviour have been discussed in some detail in \cite{hf}.

The oscillatory behaviour of the blob before relaxation, observed in this case,  brings out the relevance of the specific transport equation to be used in each situation, and the richness of the physical phenomena hidden behind  the assumption of  a vanishing relaxation time.

Let us now analyze the problem of the thermohaline convection with our kernel (\ref{1nf}). Proceeding as for the previous examples, we obtain for $DT$

\begin{equation}
\frac{d ^3DT}{d t^3}+\frac{1}{\tau}\frac{d^2DT}{d t^2}+\frac{1}{ \tau^2 \tau_d}DT(t)=0.
\label{thcc}
\end{equation}

This is a homogeneous linear, third order differential equation whose general properties are well known (see for example \cite{pp}).

In general for an equation of the form
\begin{equation}
\frac{d ^3Y}{d t^3}+a\frac{d^2Y}{d t^2}+b\frac{dY}{d t}+c Y=0,
\label{thccn}
\end{equation}
where $a, b, c$ are real constants, it can be shown that the solution is of the form (see pages 44--47 in \cite{pp})
\begin{equation}
Y=e^{\alpha t} \left(\cos{\omega t}+\sin{\omega t}\right),
\label{thin4cun}
\end{equation}
if and only if
\begin{equation}
\frac{2a^3}{27}-\frac{ab}{3}+c-\frac{2}{3\sqrt{3}}\left(\frac{a^2}{3}-b\right)^{3/2}>0.
\label{pp1}
\end{equation}
In our case $a\equiv \frac{1}{\tau}$, $b=0$ and $c\equiv \frac{1}{\tau ^2\tau_d}$, implying that the above inequality is always satisfied. The explicit form of the solution requires to solve the characteristic equation corresponding to (\ref{thcc}). This is a cubic algebraic equation whose roots are given by rather cumbersome and not very illuminating expressions at this point. Suffice is to say that  we have in this case also a damped oscillatory motion, although the  frequency and the damping  factor are different from the ones obtained with the Cattaneo--Vernotte equation.
\subsubsection{Secular stability of nuclear burning before relaxation}

The  secular stability of nuclear burning is an issue of utmost interest in astrophysics. Indeed, early work by Rosenbluth et al. \cite{acsn1} shows how  the hydrogen falling  onto the surface of a neutron star in a close binary system  undergoes nuclear fusion, independently on how low the temperature may be. This idea has been exploited by many authors for the modelling of compact x--ray sources (see \cite{acsn2,acsn3,acsn4} and references therein).

The nuclear instability reported in the above references, appears whenever the characteristic time for the increasing of the thermal energy generated by nuclear burning is smaller than the time required for the removal of this energy. 
 Also, as shown  by Hansen and Van Horn \cite{acsn2},  the range of time scales for growth of the instability span  from milliseconds
to minutes. Therefore, since the relaxation time  may be
of the order of milliseconds (or larger) for highly degenerate matter, it appears evident that the  Maxwell--Fourier law which implies  the vanishing of the  relaxation time, might not be appropriate to describe this kind of problem, and we need to  resort to transport equations including the relaxation time.

Two main ideas are required for our discussion. One is the concept of gravothermal specific heat and the other is the  equation relating the fluctuations of the energy released by nuclear burning to the variation of the temperature.

Let us start by introducing the concept of gravothermal specific heat. We shall not give much details, the reader may find a comprehensive discussion on this issue in \cite{KW}, whose approach we shall closely follow.
 
 Thus let us consider a star in hydrostatic equilibrium, whose center is surrounded by a small sphere 
of radius $r_s$ and mass $m_s$.
For a   sufficiently small  sphere,  the central values of the pressure and the density, $P_c$, $\rho_c$, may be assumed to be equal to the pressure $P$ at $r_s$ and the mean density in the
sphere, respectively. 

Let us now consider that a small amount of heat ($dq$) is added to the central sphere, producing  
its   homologous expansion, then the following expression may be found (see \cite{KW} for details):

\begin{equation}
dq = c_PT_c(\theta_c-\nabla_{ad}p_c) = c^* dT_c ,
\label{grav1}
\end{equation}
where $c_P$ is the specific heat at constant pressure, $\nabla_{ad}$ is defined by equation (\ref{nablas}), $p_c$ and $\theta_c$ are defined by:
\begin{equation}
p_c\equiv \frac{dP_c}{P_c};\qquad \theta_c\equiv\frac{dT_c}{T_c},
\label{grav2}
\end{equation}
and  $c^*$, denotes  the
gravothermal specific heat,  which relates the variations of heat and temperature, and is defined by:

\begin{equation}
c^*\equiv c_P\left(1-\nabla_{ad}\frac{4 \delta}{4\alpha -3}\right),
\label{grav3}
\end{equation}
with 
\begin{equation}
\alpha\equiv \left(\frac{\partial \ln \rho}{\partial \ln P}\right)_T;\qquad \delta\equiv -\left(\frac{\partial \ln \rho}{\partial \ln T}\right)_P,
\label{grav4}
\end{equation}
where the subscript $T$ and $P$ implies that the quantity is evaluated at constant temperature and pressure respectively.

 If  $c^*$ is positive (as  for  a nonrelativistic degenerate gas),
any increasing of energy in the central sphere would warm up the matter, which eventually 
may lead to a thermal runaway.  On the contrary if $c* < 0$ (as for an ideal monoatomic
gas), any  $dq > 0$  produces  a cooling ($dT < 0$) reducing the overproduction of
energy, avoiding thereby the thermal runaway (as it happens in the sun, fortunately!). 

So far we have not specified the source of heat. Let us now consider the case when energy is generated by nuclear reactions and transported
out of it by radiation in the diffusion approximation (assuming that the central region is not convective).

Then denoting by  $\epsilon$ and $\lambda_s$   the mean energy generation rate (per unit mass) and the energy
per unit time which leaves the sphere, respectively, the  equilibrium condition reads:
\begin{equation}
\epsilon m_s-\lambda_s=0.
\label{grav5}
\end{equation}
Next, let us now perturb (\ref{grav5}) on a time scale which is short as compared to the relaxation time $\tau$, and the thermal
adjustment time  $\tau_d$, but 
much larger than the
hydrostatic time scale.

 Then using (\ref{grav1}) and (\ref{grav5}), we obtain for the energy balance of the perturbed state :
\begin{equation}
m_sd\epsilon-d\lambda_s= m_s\frac{dq}{dt}=m_sc^*\frac{dT_c}{dt}.
\label{grav6}
\end{equation}

Next, it can be shown that, always keeping the homologous regime for simplicity, we may write (see \cite{KW} for details):
\begin{equation}
\lambda_s^{-1}d\lambda_s= 4x+4\theta_c-\kappa_p p_c-\kappa_T\theta_c,
\label{grav7}
\end{equation}
with $dr=rx$, and $\kappa_p\equiv\left(\frac{\partial \ln\kappa}{\partial \ln P}\right)_T$,  $\kappa_T\equiv\left(\frac{\partial \ln\kappa}{\partial \ln T}\right)_P$.
\\
Then, introducing (\ref{grav7}) into (\ref{grav6}) we obtain:
\begin{equation}
m_s\lambda_s^{-1}\frac{dq}{dt}= m_s\lambda_s^{-1}c^*
T_c \frac{d\theta_c}{dt}=\left[\epsilon_T+\kappa_T-4+\frac{4\delta}{4\alpha-3}(1+\epsilon_p+\kappa_p)\right] \theta_c,
\label{grav8}
\end{equation}
where $\epsilon_p\equiv\left(\frac{\partial \ln\epsilon}{\partial \ln P}\right)_T$ and   $\epsilon_T\equiv\left(\frac{\partial \ln\epsilon}{\partial \ln T}\right)_P$.
\\
The equation above  is, formally, the same independently on the transport equation, however  the explicit expression of the luminosity function $\lambda_s$ does depend on the transport equation to be used.

From a simple inspection of (\ref{grav8}), it follows that whenever  the square bracket in (\ref{grav8}),   $\lambda_s$ and    $\theta_c$  are  positive, then 
$\frac{dq}{dt}> 0$. Thus, a positive  sign of $c^*$  would produce a  heat up ($\frac{d\theta_c}{dt} > 0$) while a negative value of $c^*$ would produce 
a cool off ($\frac{d\theta_c}{dt}< 0$). It is then obvious that the explicit expression of $\lambda_s$  is critical for the final verdict about the (in)stability of the nuclear burning in any scenario.

Now, the calculation of $\lambda_s$  proceeds  as for the obtention of luminosity of the convective blob in the example analyzed in the previous subsection.
Thus we may write as in (\ref{gthhm1})
\begin{equation}
\lambda_s=S|\vec q|= \frac{2S}{d}\int^{t}_{-\infty}{Q(t-t')DT(t') dt'}.
\label{grav9}
\end{equation}
where S is the surface of the sphere and  DT denotes the temperature difference between the central sphere and the surrounding matter.

Then, the evolution equation for $DT$ is (\ref{thin4}), i.e.

\begin{equation}
\frac{d DT}{d t}=-\frac{1}{\kappa \tau_d}\int^{t}_{-\infty}{Q(t-t')DT(t') dt'}.
\label{grav10}
\end{equation}

 In the past the Maxwell--Fourier kernel has been used in (\ref{grav9}) (e.g.  \cite{KW}), leading to a  $DT$  of the form of (\ref{thin4mf}). However as mentioned before this might be inaccurate in problems where the order of magnitude of $\tau$ is larger or at least equal to the characteristic time scales of the system. That's why such calculation was performed for the Cattaneo equation in \cite{hfII}, where a $DT$ of the form given by (\ref{thin4cu}) is obtained. For our kernel (\ref{1nf}) the form of $DT$ is given by the equation (\ref{thcc}), whose solution describes a damped oscillatory behaviour  as illustrated by (\ref{thin4cun}).

In the light of previous results and comments we become aware of the relevance of processes occurring before relaxation.  Indeed, let us focus on equation (\ref{grav8}).
From standard results in nuclear physics, it can be assumed that the square bracket  in (\ref{grav8}) is positive. Therefore if  $c^* > 0$ any temperature increase would  imply  burning instability if  $\lambda_s$ is positive. But
as we have just seen, with our  kernel,  $\lambda_s$ exhibits an oscillatory behaviour,  which means   that before relaxation the system
will  jump from unstable to stable and vice-versa with a period depending on $\tau$ and $\tau_d$.
In other words, for sufficiently large relaxation time, a quasi-periodic structure in the emission is  expected to be observed before the system attains the stationary regime. Some astrophysical consequences of this fact have been discussed in detail in \cite{hfII}.
\subsection{The relativistic regime}

So far we have considered fluids in the classical regime  (non--relativistic), however  many of the applications of the subject discussed here are expected  to  be used in the study of very compact objects where Newtonian gravity is no longer reliable.  Accordingly,  it would be wise to consider the relativistic  generalization of the presented model. 

Relativistic causal dissipative theories has been the subject of many research works in the past. Particularly relevant are \cite{Is,Is2,Is3,Pa,paj1}. However, curiously enough, in the classical limit the transport equations proposed by these authors lead to the Cattaneo--Vernotte equation. Therefore it is justified to ask what is the relativistic equation producing (\ref{Cattaneo1nf}) in the non--relativistic limit.

The relativistic transport equation in the Israel--Stewart theory reads
\begin{equation}
\tau h^\mu_\nu q^\nu _{;\beta}V^\beta +q^\mu=-\kappa
h^{\mu\nu}(T_{,\nu}+T a_\nu)-\frac{1}{2}\kappa T^2\left
(\frac{\tau V^\alpha}{\kappa T^2}\right )_{;\alpha}q^\mu,\label{qT}
\end{equation}
 where $V^\beta $, $h^\mu_\nu $, $q^\mu$, $a_\nu$ denote the four--velocity, the projector on the hypersurface orthogonal to the four--velocity, the four--heat flux vector and the four--acceleration respectively. The semicolon denotes covariant derivative, and greek indices run from $0$ to $4$, with $0$ corresponding to the timelike component, whereas $1, 2, 3$ correspond to the spatial components.
 
 Let us analyze in some detail the different terms entering in (\ref{qT}), and their non--relativistic limit. For doing that we have to  keep in mind that we are considering comoving observers, implying $V^\alpha=\left(\frac{1}{\sqrt{|g_{tt}|}}, 0, 0, 0\right)$ where $g_{tt}$ denotes the $tt$ component of the metric tensor, and we are using relativistic units implying that the light velocity and the Newtonian  gravitational constant equal to one.

 The first term on the left is the ``hyperbolizer'' term, and in the non--relativistic limit becomes $\tau\frac{\partial \vec q}{\partial t}$. The second term on the left is just the heat flux vector, becoming $\vec q$ in the non--relativistic regime. Obviously these two terms must be present in any causal theory of dissipation. Next, the first term on the right of (\ref{qT}) ($\kappa h^{\mu\nu}T_{,\nu}$ ) is also trivially identified as the temperature gradient term in the non--relativistic limit, on the contrary  the second term on the right  ($T a_\nu$) is purely relativistic. It was originally discovered by Tolman \cite{Tol} when investigating the conditions of thermal equilibrium in a gravitational field, and reflects the fact that a temperature gradient is necessary to ensure the thermal equilibrium, due to the  inertia of heat. In the non--relativistic regime this term vanishes, but it must be present in any relativistic dissipative theory. 
 
 Thus, the first, the second and the fourth terms appearing in (\ref{qT}) are expected to be present in any relativistic causal theory of dissipation. Instead, the last term in (\ref{qT}) is characteristic of the Israel--Stewart theory and is absent in  some other theories, (e.g. \cite{PAN}). It should be mentioned that in some relativistic theories  additional terms  may be added which vanish in the classical limit. 
 
 Therefore, as mentioned before, the non--relativistic limit of (\ref{qT}) is just the Cattaneo--Vernotte equation, and we have to propose something different to (\ref{qT}) in order to recover (\ref{Cattaneo1onf}) in the non--relativistic regime.
 
 From the above comments we see that the options are restricted to changes in the third term. The simplest modification of  (\ref{qT}) leading to (\ref{Cattaneo1onf}) in the non--relativistic limit, reads:
 
\begin{equation}
\tau h^\mu_\nu q^\nu _{;\beta}V^\beta +q^\mu= -\kappa h^{\mu\nu} \int^{t}_{-\infty}T_{,\nu} \sqrt{|g_{tt}|}dt'
-\kappa T a^\mu-\frac{1}{2}\kappa T^2\left
(\frac{\tau V^\alpha}{\kappa T^2}\right )_{;\alpha}q^\mu.
\label{qTm}
\end{equation}

It goes without saying that, for any specific problem, using (\ref{qTm}) instead of (\ref{qT}) would lead to quite different conclusions.  However any of such application is out of the scope of this work and we only wish to stress here  the relevance of this point.

\section{Conclusions}
We have put forward a new model of heat causal conduction not satisfying the fading memory paradigm. Although we are well aware of the fact that an almost infinite set of kernels violating the fading memory paradigm may be conceived, we wished here to propose a specific kernel in order to bring out the effects of this type of kernel in the two examples analyzed. Besides, it is worth mentioning that  our kernel somehow represents the extreme case of violation of the fading memory paradigm.

 Two specific applications were discussed in some detail to illustrate the critical dependence of the physical analysis of each scenario on the specific transport equation involved. The presented analysis clearly  exhibits  the observational differences appearing as the result of using our kernel instead of Maxwell--Fourier or Cattaneo. However much more detailed setups are required to propose specific experiments to confront those alternatives.
In the same line of arguments it would be interesting to study different solutions to (\ref{2}) under a variety of circumstances and compare the resulting temperature profiles with those obtained from the integration of (\ref{telegraph}).

For the sake of completeness we have also proposed a simple generalization of our model to the relativistic regime. Further applications of the resulting equation to specific problems involving relativistic dissipative processes are required to deduce the possible observational implications of the model.

The problem under consideration here may be approached by means of  different frameworks, an example of which is the rational extended thermodynamics. It would be interesting to explore if new aspects of the proposed model could be brought out using any of these theories.

Finally, it is worth dedicating some thoughts about the possible situations where we could expect a dissipative process not complying with the fading memory paradigm. Obviously, for sufficiently large relaxation times the impact of the temperature gradients  at the neighborhood of the observation time, on the observed heat flux, may be negligible. In this sense our kernel (\ref{1nf}), as mentioned above,  represents an extreme situation. Constitutive equations for the internal energy  different from (\ref{ene}), could also lead to the same kind of kernel.

\section{Acknowledgements}
This work was partially
supported by the Spanish Ministry of Science and Innovation (grant
FIS2010-15492).

\end{document}